**Inexpensive electronics and software for photon statistics and correlation spectroscopy**

Benjamin D. Gamari,[1] Dianwen Zhang,[1,a] Richard E. Buckman,[1] Peker Milas,[1]

John S. Denker, Hui Chen,[2] Hongmin Li,[2,3] and Lori S. Goldner[1,b]

[1]*Department of Physics, University of Massachusetts, Amherst MA 01002*

[2] *Wadsworth Center, New York State Department of Health, 120 New Scotland Ave, Albany NY 12208*

[3] *Department of Biomedical Sciences, School of Public Health, State University of New York at Albany, PO BOX 509, Albany NY 12201-0509*

Single-molecule sensitive microscopies and spectroscopies are transforming biophysics and materials science laboratories. Techniques such as fluorescence correlation spectroscopy (FCS) and single-molecule sensitive fluorescence resonance energy transfer (FRET) are now commonly available in research laboratories but are as yet infrequently available in teaching laboratories. We describe inexpensive electronics and open-source software that bridges this gap, making state-of-the-art measurement research capabilities accessible to undergraduates interested in biophysics. We include a pedagogical discussion of the intensity correlation function relevant to FCS and its calculation directly from photon arrival times. We demonstrate the system with a measurement of the hydrodynamic radius of a protein using FCS that is suitable for an undergraduate teaching laboratory.  The FPGA-based electronics, which are easy to construct, are suitable for more advanced measurements as well, and several applications are demonstrated. As implemented, the system has 8 ns timing resolution, outputs to control up to four laser sources, and inputs for as many as four photon-counting detectors.

[a] Current Address: Beckman Institute for Advanced Science and Technology, University of Illinois at Urbana-Champaign, 405 North Mathews, Urbana IL 61801 USA

[b] Author to whom correspondence should be addressed. Electronic mail: lgoldner@physics.umass.edu

## I. INTRODUCTION

Through techniques such as fluorescence correlation spectroscopy (FCS) and dynamic light scattering (DLS), photon statistics have provided a window into molecular and materials properties for over 40 years. Common to these widely-used techniques is the need to rapidly and efficiently collect photons and calculate photon correlation functions with time resolution below 1 μs. In recent years, high quantum-yield, low dark count photodetectors have become commonplace, lowering the detection limit to a level that makes it possible to routinely observe fluorescence from individual molecules. At the same time, the availability of fast timing circuitry has meant that the arrival time of every detected photon can be determined with high resolution.[1,2]  With this detailed information about the photon stream has come many new opportunities for examining photon statistics.[1,2] Single-molecule-sensitive fluorescence measurement, which is now revolutionizing our understanding of molecular biophysics, has particularly benefitted from better hardware.  High-end commercial instruments[1-3] and software[1,4,5] for evaluating photon statistics are commonly found in biophysics, materials research, and optics laboratories; however the cost of these systems, which range from a few thousand to tens of thousands of dollars depending on functionality, is beyond the reach of nearly all teaching laboratories and many research laboratories.

Our intent is to provide inexpensive electronics and associated open-source software that brings state-of-the-art capabilities for measuring and analyzing photon statistics into the teaching lab while providing a valuable lesson in instrumentation for intrepid students. We include a pedagogical discussion of the intensity correlation function relevant to FCS and its calculation directly from the photon arrival times that constitute the data in most modern instruments. Graduate students or other researchers interested in building or better understanding their own hardware and software may also find this work useful.

The examples and applications used here all involve fluorescence detection. The use of fluorescent dyes to study biological or organic materials is very common, and many students have at least some experience with staining. Staining often consists of the random intercalation of dyes and is useful mostly for visualization. In molecular biophysics, it is common to modify a specific biomolecule by covalent attachment of a fluorophore (dye molecule). In that case, to the extent that the fluorophore's photophysical properties are coupled to the physical motion (including linear and rotational diffusion, folding, twisting and bending), or chemical kinetics (binding) of the host biomolecule, the dye can be used as a nanoscopic reporter of these molecular properties. The photons emitted from the fluorophore carry information about the biomolecule that is received by a detector and timing hardware, and decoded by statistical analysis.

As correlation functions often provide a starting point for a discussion of photon statistics, fluorescence correlation spectroscopy (FCS) is discussed in detail here. In FCS, fluctuations in dye brightness that are coupled to molecular folding or binding might be used to infer the correlation or relaxation times for these processes.[6] However, much of the utility of the technique in biophysics arises from the relationship between diffusivity, $D$, and molecular size given by the Stokes-Einstein relation:[7]

$$D = \frac{k_B T}{6\pi\eta R_h} \qquad (1)$$

where $k_B$ is Boltzman's constant, $T$ is the absolute temperature, $\eta$ is the dynamic viscosity of the medium, and $R_h$ is the hydrodynamic radius of the diffusing particle. In FCS, fluorescent molecules are permitted to diffuse freely through the detection volume ($\cong 1$ fL) of a confocal microscope. The diffusion time through the volume is determined by the Brownian motion of the particle, which in turn depends on the diffusivity. If the concentration of the fluorophores is sufficiently low, then the largest fluorescence intensity fluctuations occur simply because the average number of molecules in the detection volume fluctuates substantially as molecules diffuse in and out. The typical diffusion time for a molecule to cross the detection volume in a plane perpendicular to the optical axis of the microscope is given by $\tau_D = \left\langle \rho^2 \right\rangle \big/ 4D$, where $\rho$ is the radial distance from the center of a cylindrically symmetric detection volume. This diffusion time is reflected in the correlation function of the fluorescence intensity (Eq. (2) and Sec. III) as a characteristic decay time. A determination of the diffusion time using FCS gives $D$ and therefore also $R_h$ using Eq. (1); a specific example is given in Sec. IV.A. We note that the use of FCS to observe Brownian diffusion and measure equilibrium constants using a simple instrument was the topic of a previous article in this journal,[8] which also covered the complementary topics of modeling and fitting to the correlation function data.

Until about ten years ago, photon correlation functions were usually generated by hardware designed specifically for this purpose. This process is efficient but costly in terms of information, since no representation of the original photon stream is stored. Alternately, a record of the intensity as function of time might be stored as photon counts detected in sequential time bins; this approach is used in Ref. 8. For photons detected in microsecond or faster time bins, this intensity data can be very sparse. In FCS, it is not unusual for <u>average</u> photon count rates to be a few kHz; even at 1 μs resolution, the vast majority of bins will be empty. The more recent alternative, discussed here, is to record directly the arrival time of each photon. This method has the distinct advantage of providing the most information about the photon stream in a compact manner. In this new data acquisition paradigm, the correlation function, given in FCS by:

$$G(\tau) = \frac{\langle I_a(t) I_b(t+\tau) \rangle}{\langle I_a(t) \rangle \langle I_b(t) \rangle} \tag{2}$$

is most efficiently calculated directly from the arrival times.  Here $I$ denotes fluorescence intensity and the subscripts $a$ and $b$ can denote data taken in a single channel ($a = b$, autocorrelation) or in different detector channels, corresponding to different colors or polarizations (cross-correlation). While prior discussions of how to calculate correlation functions from photon arrival times are extant in the literature,[4] none are accessible at the undergraduate level.

 Examples of more advanced biophysical measurements using this system are given in Secs. IV.B and C.   In fluorescence resonance energy transfer (FRET), which is described in Sec. IV.B, two different dyes, rather than a single dye, are attached to a biomolecule. Changes in distance between the dye molecules are recorded as changes in their relative fluorescent intensities, the result of a resonant dipole-dipole interaction. By combining FRET with FCS,[9] the timescale and nature of fluctuations in molecular structure occurring on a nanometer length scale can be determined with time resolution limited only by the electronics.

To facilitate fast, multi-channel detection, as well as control of multiple sources for fluorescence excitation (*e.g.*, different wavelength lasers lines to excite different dyes), the electronics are based on a field-programmable gate array (FPGA). Using the FPGA, the electronics are straightforward to construct (discussed in Sec. II and the Supplementary Materials) and provide an opportunity to explore instrumentation that makes a good bridge for students interested in experimental research. The instrument described has eight input channels, four sequencer outputs for control of excitation light, and 8 ns timing resolution with no dead time between events. The use of an FPGA ensures design flexibility; notably, the number and configuration of the input and output channels are easily modified to suit different experimental requirements. The FPGA used here costs under $200; with the required oscillator, connectors, and break-out box the total cost should be under $400. Software and firmware necessary to implement and run the device are included in the Supplementary Materials.

It is worth noting that high-end systems (*e.g.*, Picoquant or Becker & Hickl) that are designed for more demanding time-correlated single-photon counting applications have timing accuracy measured in picoseconds and cost measured in tens of thousands of dollars. Home-built systems based on a data acquisition card (*e.g.*, the National Instruments PCI 6602) can be constructed for about $2k but have only microsecond resolution.[10] A card with 42 ns timing resolution and four input channels is sold by ISS for about $15k. None of these systems incorporate output channels for laser control.

## II. INSTRUMENTATION

**A. Optics.** The minimal optical instrumentation necessary for FCS consists of a confocal microscope, laser, filters, and detector. Rieger *et al.*[8] give a detailed description, including parts list, for a home-built FCS microscope suitable for use in teaching that includes laser and detector and that costs approximately $6k. Alternately, an open-frame FCS system kit suitable for both teaching and research applications is currently under development through collaboration with Thorlabs (part number TFCSK) and will be available for about $8k; with a DPSS laser and detector, this system might cost $10.5k to assemble.

The optics used here are described in Fig. 1. The excitation light is coupled into the microscope via a single-mode optical fiber, which also serves as a spatial filter in place of the pinhole in Rieger *et al.*[8] An acousto-optic tunable filter (AOTF, NEOS part number 48058-5-.55) is used to switch between excitation lines but need not be included in a teaching apparatus that uses a single laser line. A narrow-pass single- or dual-band excitation filter is necessary before the dichroic mirror to eliminate leakage from unwanted laser lines or the DPSS pump; the dichroic mirror reflects the excitation light and transmits longer wavelength fluorescence. The telescope after the fiber serves two purposes. First, it works to expand the laser beam diameter to match the size of the back aperture of the objective lens. This ensures that the entire numerical aperture of the lens is used in focusing the laser beam, thereby minimizing the corresponding focal volume. The telescope also serves to image a steering mirror to the back focal plane of the objective, greatly simplifying alignment of the microscope. The objective used here is an Olympus UPlanSApo 60X with a numerical aperture (NA) of 1.2. A high NA water or oil objective is necessary for FCS. The diffraction-limited focal volume of these lenses is typically 1 fL, leading to a similarly small and well-defined detection volume that satisfies the low-background requirements of FCS or single-fluorophore detection.

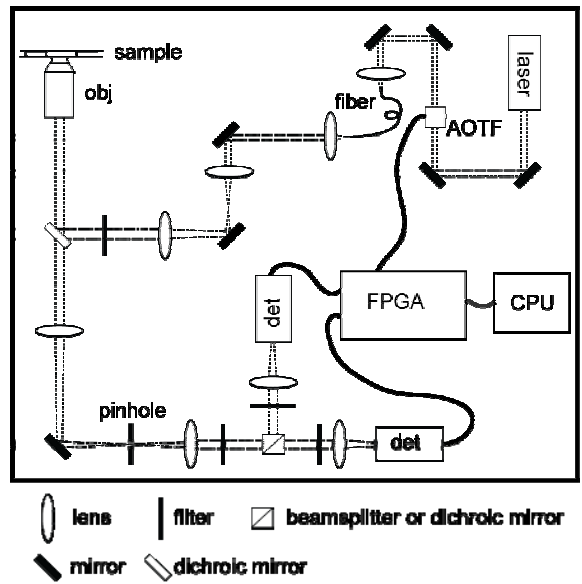

FIG. 1. Schematic of the optics. Obj, objective; AOTF, acousto-optical tunable filter; FPGA, field programmable gate array; det, detector. The photon counting detectors output TTL level pulses whose arrival times are recorded by the FPGA. The FPGA also controls the operation of the AOTF and registers the timing of its state changes.

Fluorescence from the focal (detection) volume of the microscope is collected back through the objective and imaged through a 50 μm diameter pinhole. Pinhole size depends on choice of objective, beam diameter, and focusing lens (tube lens) and is roughly chosen to match the size of the focal spot.[11] The tube lens under the dichroic should be a good achromat to avoid aberrations at the pinhole. Alignment of the instrument with a single detector is discussed in detail in Ref. 8.

In this work, two detectors are in use. In fluorescence correlation spectroscopy (FCS), it is often convenient to use a non-polarizing beamsplitter between two detectors so that the instrument is arranged in a Hanbury Brown and Twiss configuration (HBT),[12] with the two detectors sampling the same signal. In this configuration, the cross-correlation is insensitive to detector deadtime and afterpulsing.[13] For example, with a single channel measurement it is not possible to observe correlations below the deadtime of the detectors (typically 100 ns), and below 1 μs the signal is dominated by detector afterpulsing. In the HBT configuration, photon timing between channels is limited only by the speed of the electronics, which have no intrinsic deadtime.[2]

Alternately, the detectors can be set up for polarization and/or two-color measurements by replacing the non-polarizing beamsplitter with a polarizing beamsplitter or dichroic mirror. In Sec. IV, a dichroic mirror is in place between the detectors. Detectors (det) are high-efficiency, low dark count, photon-counting modules such as Hamamatsu's model H7421-40 photomultiplier tube, or Picoquant's τ-SPAD single photon-counting avalanche photodiode. A Cheaper alternative suitable for a teaching laboratory is the Hamamatsu H10682-110 (approximately $1300). All of these units output TTL level pulses.

**B. Electronics.** The electronics are diagrammed in Fig. 2. The device is built on the KNJN Xylo EM FPGA development board. This board includes an Altera Cyclone II (EP2C5) FPGA connected to the host computer through a Cypress FX2 USB interface.

No additional hardware or modifications to the Xylo are necessary, although an external crystal oscillator at 32 MHz was installed and multiplied on chip to 128 MHz to achieve timing resolution of 7.8 ns. An overview of operation is given here; detailed construction information can be found in the Supplementary Materials.

The hardware records the arrival time of digital events (*e.g.*, TTL level pulses from the photon-counting devices) occurring on up to four channels, and it also provides four digital outputs for experimental control. Here the outputs control the acousto-optic tunable filter (AOTF) that is used to switch between excitation wavelengths; with single-line excitation typical in a teaching laboratory the output lines are unused.

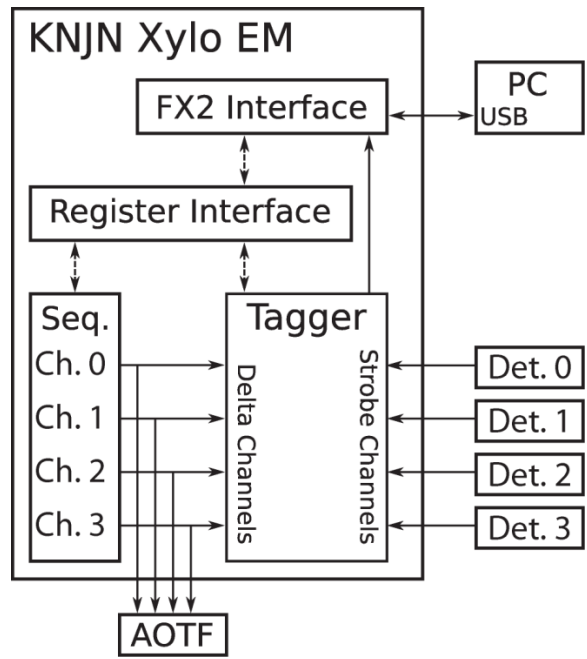

FIG. 2. Schematic of electronics. The device is built on the KNJN Xylo EM FPGA development board. This board includes an Altera Cyclone II (EP2C5) FPGA connected to the host computer through a Cypress FX2 USB interface. The Tagger records the arrival time of photons from up to four detectors on the Strobe channels, and sequencer (Seq) events on the delta channels. The sequencer controls laser-line switching, if needed, through the AOTF.

The FPGA's firmware[14] is derived from the time-tagging multiple-coincidence detector design developed by Polyakov *et al.*[15] The code was adapted for better extensibility, and extended to accommodate additional input channels and digital outputs for experimental control and timing. The firmware is split into two logical blocks: the time tagger and the sequencer.

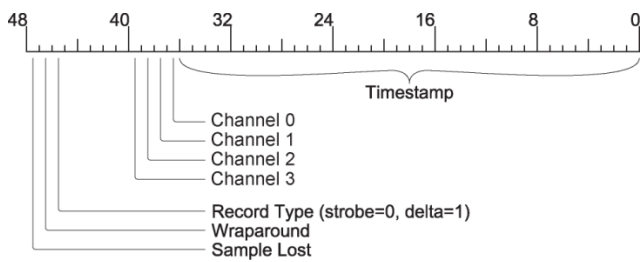

FIG. 3. The data record format. The first 36 bits contain the event time in clock cycles. The latter part of the record contains ancillary information about the event. Bits 37-40 indicate the event's originating channel (or channels) and bit 46 indicates the type of event (strobe or delta). Bits 47 and 48 indicate counter overflows and dropped record events, respectively.

The time tagger (Tagger in Fig. 2) is responsible for recording the arrival time of events on a set of inputs. The inputs are denoted "delta" and "strobe." Delta inputs record changes of the digital output channels (sequencer events), while strobe inputs record photon timing signals (photon events). Strobe channel inputs are TTL level signals generated by the photon-counting detectors. Delta channel inputs are TTL level signals generated by the sequencer outputs. Strobe and delta data types are recorded in the same format by the Tagger. This record format is shown in Fig. 3. The first 36 bits contain the event time in clock cycles, and the latter part of the record contains ancillary information about the event. For strobe events, bits 37-40 indicate the event's originating channel (or channels). For delta events, bits 37-40 record the state of the sequencer immediately after a change. Bit 46 indicates the type of event (strobe or delta). Bits 47 and 48 indicate counter overflows and dropped record events, respectively.

The sequencer module is designed to control laser excitation sources. Each sequencer channel can be programmed to produce a periodic TTL waveform. Sequencer outputs can be used, *e.g.*, to drive the digital input channels of an AOTF, thereby controlling the wavelength of the excitation light. Changes of the sequencer outputs, and therefore the excitation wavelength(s), are recorded at the delta channel inputs of the event tagger as described above. In this way, output state transitions of the sequencer are recorded alongside photon events recorded at the strobe channels. This facilitates the implementation of alternating laser excitation in Sec. IV.C, and with different electro-optic elements might also be used in polarization switching or photoactivation.

Software and firmware packages to program and run the electronics, and installation instructions, are included in the Supplementary Materials. Written for a Linux platform, they include the necessary firmware (in the packages called "timetag-fpga" and "timetag-fx2"), a user interface for experimental control ("timetag-tools"), and data manipulation and analysis tools written in Python and C++ ("photon-tools").

### III. CALCULATION OF THE CORRELATION FUNCTION FOR FCS

**A. Conceptual overview.** The correlation function used in FCS is given by Eq. (2). The intensity $I(t)$ is not usually a continuous function of time, but rather consists of discrete photon arrival events. The <u>average</u> photon count rates in FCS typically range from 1 kHz to 100 kHz. However, the timescale of interest in most measurements is frequently well below 1 ms, and data are acquired with a timing accuracy 1 ps $\leq \Delta t \leq$ 1 µs. This implies a sparsity to the intensity data that makes conventional numerical schemes for storing data and performing calculations inefficient. Conventionally, photon counts might be collected for a time $T$ in time bins of length $\Delta t$. For sparse data, the number of bins required, $(T/\Delta t)$, can far exceed the total number of photons acquired, which is typically of order $10^6 - 10^8$. For this reason it is often more efficient, as well as more informative, to record photon arrival times directly.

Nonetheless, we begin our conceptual discussion with a conventional discretization of $I(t)$ into an array $I(t_i)$ with $T/\Delta t$ elements. The numerator in Eq. (2) can be considered a "lagged dot product." The meaning of this can be understood with the help of Fig. 4, which illustrates the case where channel $a$ and channel $b$ each have 20 elements (*e.g.,* bins). Each matrix element (small square) represents the product $I_a(t_i) I_b(t_j)$. The numerator of Eq. 2 at $\tau = 0$ is represented by summing over the main diagonal in the diagram, for which $j = i$. Similarly, at $\tau = 4$ the numerator is represented by summing along a path parallel to the main diagonal for which $j = i + 4$, as shown in the figure.

Choice of boundary conditions is also illustrated in Fig. 4. We consider three possibilities: periodic boundaries, "dark" boundaries, and "extended" boundaries. In the case of periodic boundaries with 20 elements in each channel (Fig. 4), the evaluation of the numerator in Eq. (2) for a lag of $\tau = 4$ includes both the path at $\tau = 4$ and also the path labeled $\tau = -16$. When using periodic boundaries, there will always be $T/\Delta t$ squares contributing to the lagged dot product. This simplifies the calculation of the denominator in Eq. (2), since it need be calculated only once for a particular choice of bin size. However, $G(\tau)$ calculated with periodic boundaries can sometimes yield unphysical results if $I(t)$ is not periodic and has correlations at timescales that are not very much less than $T$. Periodic boundaries are implicit in the use of a Fourier transform to calculate the correlation function.

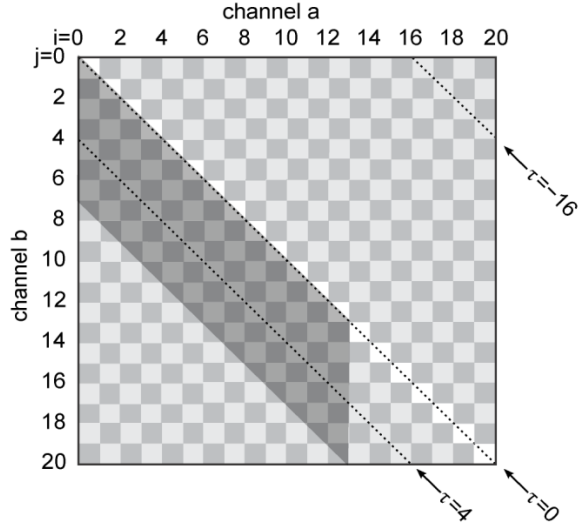

FIG. 4. Graphical example of the calculation of the lagged dot product in the numerator of Eq. (2) with various boundary conditions. Index $i$ ($j$) is used to denote specific bins in channel $a$ ($b$). Each square should be understood to contain the product $I_a(t_i) I_b(t_j)$. For dark boundary conditions, the dot product at a given lag is represented by the sum of the matrix elements along a line parallel to the main diagonal. With 20 bins per channel as shown here, periodic boundary conditions result in the matrix elements at $\tau = -16$ contributing the dot product at $\tau = 4$. Extended boundary conditions use only those matrix elements inside the shaded region.

For dark boundary conditions, we include only the path labeled $\tau = 4$ in the calculation of the lagged dot product. As a consequence, the number of contributing squares decreases as the lag increases. The denominator in Eq. (2) should reflect averages over only that part of each intensity that is used in the corresponding numerator, which changes at every $\tau$. That is, the same subset of data that contributes to the numerator in Eq. (2) also contributes to the denominator. In some treatments, the denominator in Eq. (2) is written $\langle I_a(t) \rangle \langle I_b(t+\tau) \rangle$ to indicate this, but this still misses the fact that $\langle I_a(t) \rangle$ is calculated over the range $0 < t_i < T - \tau$ while $\langle I_b(t+\tau) \rangle$ is calculated over the range $\tau < t_i < T$.

For extended boundary conditions, only the data inside of the shaded parallelogram in Fig. 4 is used. The same number of squares, and approximately the same number of photons, are used in each calculation of $G(\tau)$. Data in channel $a$ with $t_i > T - \tau_{\max}$ will not be used at all, where $\tau_{\max}$ is the maximum lag time to be calculated. The subset of data in channel $b$ that contributes to the correlation function changes with each lag time: only data in the window between with $\tau < t_j < T + \tau - \tau_{\max}$ will contribute. The denominator in Eq. (2) does not change systematically with $\tau$, but will fluctuate as the data contributing to the numerator changes. In a properly designed experiment, $T \gg \tau_{max}$ so only a small amount of data is unused. Extended boundaries are implemented here.

From Fig. 4 it can be seen that the calculation of the numerator of Eq. (2) requires approximately $T/\Delta t$ calculations at each lag $\tau << T$. Calculating $G(\tau)$ for all $\tau$ with fixed resolution $\Delta t$ therefore scales like $(T/\Delta t)^2$. This can be reduced considerably using a "multi-tau" algorithm,[16] initially introduced for hardware correlators, for reducing the number of lag times $\tau$ at which to calculate $G(\tau)$. As the spacing between lag times increases, the bin width is also increased, leading to a calculation that requires only $(T/\Delta t)\log(T/\Delta t)$ operations. [16] It is worth noting that a calculation of $G(\tau)$ from the Fourier transform of the power spectrum[17] also scales with $(T/\Delta t)\log(T/\Delta t)$ but maintains the original resolution $\Delta t$ between successive $\tau$.

An example of rebinning necessary to implement a multi-tau algorithm is diagrammed in Fig. 5. Each square should be understood to contain the product $I_a(t_i)\, I_b(t_j)$. As usual, the correlation function at a given lag is represented by the sum of the matrix elements along a line parallel to the diagonal, with $G(0)$ given by a sum along the diagonal, subject to the chosen boundary conditions. The first eight computations, $0 \leq \tau \leq 7$, use a bin size $\Delta t$. Then, starting at $\tau = 8$, the bin size is doubled. By the overlap of small and large squares at $\tau = 8$, it is evident that some of the photons that contributed to the sum for $\tau = 7$ also contribute to $\tau = 8$. This leads to some small amount of smoothing of the correlation function at the rebinning boundaries, and will contribute to higher order correlations that can complicate the analysis of uncertainties. For these reasons, rebinning is generally done only once an octave. In this example, at $\tau = 16$ the bin size is again doubled to $4\Delta t$.

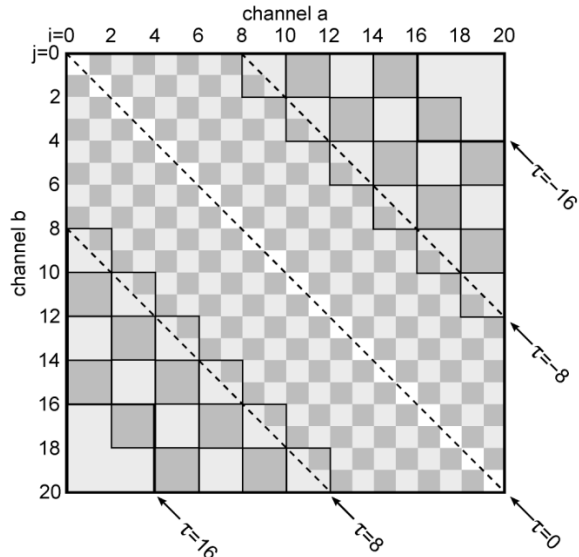

FIG. 5. Graphical representation of the calculation of the lagged dot product with rebinning for conventional, sequentially-binned lists. Each square should be understood to contain the product $I_a(t_i)\, I_b(t_j)$. The numerator of Eq. (2) at a given lag is represented by the sum of the matrix elements along a line parallel to the diagonal, with $G(0)$ given by a sum along the diagonal itself. Referring to Sec. III.B, in this example $\Gamma = 4$ and $\ell = 1$. In the first stage of the calculation, $0 \leq \tau \leq 7$. Starting at $\tau = 8$, the bin size is $\ell = 2$. At $\tau = 16$ the bin size doubles again. The small number of bins and relatively large lag times are used to simplify this example; in a well-designed experiment $T \gg \tau_{max}$ and only a small portion of the matrix around the diagonal contributes to the signal.

**B. Calculation from time tags.** It is possible to calculate $G(\tau)$ more efficiently by recognizing and utilizing the sparse nature of the data.[4] Consider a list of photon arrival times (time tags) and associated photon counts $\{(t_1, \mu_1), (t_2, \mu_2), \ldots (t_\alpha, \mu_\alpha)\}$, which as a shorthand we write as the list $\{(t, \mu)\}$ with $\alpha$ entries. The arrival times $t_i$ are measured with precision $\ell$ that must be a multiple of the hardware clock time $\Delta t$ (for raw data, $\ell = \Delta t$). In this compact representation of the data, the photon count $\mu_i$ is always greater than or equal to one. In other words, bins with zero photons are omitted from the list. For very sparse data, $\mu_i = 1$ and the number of entries $\alpha = M$, where $M$ is the number of photons represented by the list. For dense data, the $t_i$ are sequential, $\mu_i$ is just the number of photons in the $i^{\text{th}}$ bin, and the number of entries in the list is the same as the number of bins, $\alpha = N \equiv T/\ell$. If we now consider two such lists $\{(t^a, \mu^a)\}$ and $\{(t^b, \mu^b)\}$, with $\alpha$ and $\beta$ entries respectively, Eq. (2) can be approximated as:

$$G(\tau) \cong \frac{\frac{1}{N}\sum_{i,j} \mu_i^a \mu_j^b \delta_{t_i, \tau + t_j}}{\frac{M_a}{N}\frac{M_b}{N}} \quad , \tag{3}$$

where $M_a = \sum_i \mu_i^a$ and $M_b = \sum_j \mu_j^b$ are the total number of photons in each list, and $\delta$ is the Kronecker delta. The limits on the sum over the *i, j* depend on choice of boundary conditions, as discussed above. We have assumed that the precision with which the time tags are measured is the same in both lists. For sparse data, each computation of this sum will require $M_a + M_b$ calculations (*i.e.,* the total number of photons), which can be many orders of magnitude smaller than $T/\Delta t$ .

The sum in the numerator of Eq. (3) is just the lagged dot product of the discretized intensities expressed in a compact form. Our implementation of this compact lagged dot product is described by the pseudo-code in Fig. 6.

```
DotProduct({(t^a, μ^a)}, {(t^b, μ^b)}, τ_ℓ)
────────────────────────────────────────────
i ← 1; j ← 1; dot ← 0
// α and β are the lengths of {(t^a, μ^a)} and
{(t^b, μ^b)}
while i ≤ α and j ≤ β do
        diff ← t_i^a − (t_j^b + τ_ℓ)
        if diff = 0 then
                dot ← dot + μ_i^a · μ_j^b
        if (diff ≤ 0) then i ← i + 1
        if (diff ≥ 0) then j ← j + 1
return dot
```

FIG. 6. Psuedo-code representing the lagged dot product using a compact representation of the data, $\{(t, \mu)\}$, where the $t$'s are a list of time tags and the $\mu$'s, which are always $\geq 1$, are the number of photons arriving in a bin at time $t$. Here the list $\{(t^a, \mu^a)\}$ has $\alpha$ elements, and the list $\{(t^b, \mu^b)\}$ has $\beta$ elements. $\tau_\ell$ is a lag time measured in units of the bin width, $\ell$. $\leftarrow$ is an assignment operator and is often read as "gets".

To implement multi-tau analysis,[16] the spacing between lags ($\Delta\tau$) increases in stages with $\tau$, while $\ell$ (the timing precision or bin size) also increases, thus decreasing $N$. The bin size $\ell$ is chosen to be sufficiently smaller than $\tau$ to obviate the much-discussed triangular error in the approximation of Eq. (2).[18] Our implementation of the rounding (*i.e.,* rebinning) step for a multi-tau analysis is described by the pseudo-code of Fig. 7.

```
Rebin({(t, μ)})          // Input bins of size ℓ
────────────────────────────────────────────────
(t'_1, μ'_1) ← (⌊t_1 / 2⌋, μ_1)
j ← 2
// α is the length of {(t, μ)}
for i = 2 to α do
        if t'_{j−1} = ⌊t_i / 2⌋ then
                (t'_{j−1}, μ'_{j−1}) ← (t'_{j−1}, μ'_{j−1} + μ_i)
        else
                (t'_j, μ'_j) ← (⌊t_i / 2⌋, μ_i)
                j ← j + 1
return {(t', μ')}      // Output bins of size 2ℓ
```

FIG. 7. Pseudo-code for changing the bin size, which is the same as the resolution of the time tags $t$, by a factor of two. $\leftarrow$ is an assignment operator and is often read as "gets". $\lfloor \rfloor$ means "lower bound". Since the $t$ are integers, division by two followed by the lower bound operator effectively increases the bin size, and decreases the number of bins, by a factor of two.

The complete calculation of $G(\tau)$ is outlined by the pseudo-code in Fig. 8. The calculation is organized in a series of stages. The minimal time resolution or bin size $\ell = \Delta t$ is used only for the first 16 possible lag times, where $0 \leq \tau \leq 15\Delta t$. Thereafter, the bin size is increased by a factor of two at the beginning of each stage; steps in $\Delta \tau$ are also increased by a factor of two at the beginning of each stage. After the first 16 lags, a new stage begins at every doubling of the lag time, *i.e.*, every octave in $\tau$. In each stage except the first, $\Gamma$ different equally-spaced values of $\tau$ are evaluated; by default, $\Gamma = 8$. In the pseudo-code example, after the first 16 lags have been evaluated, the bin size is doubled for the next 8 lag times, $16\Delta t \leq \tau \leq 30\Delta t$ ($8 \leq \tau^g \leq 15$ where $\tau^g$ is expressed in units of the new bin size of $\ell = 2\Delta t$). The bin size continues to double with each octave in $\tau$.

---

Correlation($\{(t^a, \mu^a)\}, \{(t^b, \mu^b)\}, \tau_{max}$)

$\Gamma \leftarrow 8$ // Constant: calculations per octave
$\ell \leftarrow \Delta t$ // bin size
$\tau_\ell \leftarrow 0$ // $\tau$ measured in bins
**while** $\ell \tau_\ell \leq \tau_{max}$ **do**
 **if** $\tau_\ell = 2\Gamma$ **then**
  $\{(t^a, \mu^a)\} \leftarrow$ **Rebin**($\{(t^a, \mu^a)\}$)
  $\{(t^b, \mu^b)\} \leftarrow$ **Rebin**($\{(t^b, \mu^b)\}$)
  $\ell \leftarrow 2\ell$ // resize bins
  $\tau_\ell \leftarrow \Gamma$ // set $\tau_\ell$ to octave start
 $G(\ell \tau_\ell) \leftarrow \dfrac{\textbf{DotProduct}(\{(t^a, \mu^a)\}, \{(t^b, \mu^b)\}, \tau_\ell)}{N \cdot (M_a/N) \cdot (M_b/N)}$
 $\tau_\ell \leftarrow \tau_\ell + 1$
**return** G

---

FIG. 8. Pseudo-code for calculation of Eq. (3) with rebinning as described in the text. $\tau$ is a lag time measured in units of the bin width, $\ell$. $\leftarrow$ is an assignment operator and is often read as "gets". After the first stage of the calculation, $G(\tau)$ is evaluated at $\Gamma$ different equally-spaced values of the lag time $\tau$, ranging from $(\Gamma) \tau_\ell$ to $(2\Gamma - 1) \tau_\ell$, for each octave in the lag time. Here $N \equiv T/\ell$, and $M_a$ and $M_b$ are the total number of photons in each list. $T$ is the total time of the measurement, which is presumed to be the same in both channels.

The denominator in Eq. (3) is given by the multiplication of the total number of photons in each compact list ($M_a M_b$) divided by $N^2$. The total measurement time of the experiment, $T$, is usually the same in both channels, so that $N = T/\ell$ is always the same in both channels. At a given value of the bin size $\ell$, and for most experimental conditions (*i.e.*, $T \gg \tau_{max}$ and the system in equilibrium), the denominator is will change only slightly at each $\tau$ (if at all) in a way that depends on the boundary conditions.

**C. Calculation of uncertainty in the correlation function.** The most straightforward method to find the experimental uncertainty in the correlation function is by direct calculation of the standard deviation of the mean of several measurements.[19] Typically, a single data run might contain $10^6$ - $10^7$ photons. Dividing the data into $Q$ temporal subsets (typically 10), we calculate:

$$\sigma(\tau) = \frac{\sqrt{\left\langle G(\tau)^2 \right\rangle - \left\langle G(\tau) \right\rangle^2}}{\sqrt{Q-1}} \tag{4}$$

This works so long as $G(\tau) \to 1$ at long lag times and the sample is in equilibrium, *i.e.*, there is no drift in the signal from data subset to data subset.

The statistical uncertainty and bias in $G(\tau)$ has been reviewed and discussed in detail by Saffarian and Elson,[20] who base much of their discussion on earlier work by Schatzel.[18,21] A good understanding of the uncertainty in correlation data requires a model of the underlying photophysics of the signal. The calculation of the uncertainty by Eq. (4) does not tell the complete story of uncertainty in the correlation function; in particular, the calculation of the standard deviation alone does not describe the strongly correlated noise at different lag times that is to be expected at short lags.[22]

# IV. EXAMPLES

In this section we describe several measurements performed using the hardware and software described.  The first of these, like the examples in Rieger *et al.*,[8] is suitable for use in a teaching laboratory.  The remainder of the examples are more advanced to implement, but serve as an introduction to the many other biophysical measurement techniques that are possible using this system.  Examples IV.B-IV.D may also be of particular interest to graduate students or researchers interested in implementing this low-cost system in their laboratories.

**A. Measuring the hydrodynamic radius of a fluorescent protein.**  Fluorescent proteins, which were first isolated from the jellyfish *Aequorea victoria*, have become one of the most ubiquitous probes of protein expression, localization, and transport in living cells.[23] Yellow Fluorescent Protein (YFP, molecular mass 27 kDa) can be purchased in lyophilized form from Biovision Inc.  Alternately, it can be expressed using the plasmid pEYFP-C1 available through Clontech.

YFP was prepared to a final concentration of 5.0 nmol/L protein in 50 mmol/L glycine (pH =7) and 0.16 mmol/L Tween 20 detergent. The detergent helps prevent protein aggregation and surface adsorption.  Preparation and loading of a sample cell coated with bovine serum albumin (BSA) is described by Rieger *et al.*[8] The sample is placed on the microscope stage and the focus is positioned approximately 30 microns above the coverslip surface.  YFP was excited with 10 μW (measured before the objective) from the 514 nm line of the Argon-Krypton laser. Fluorescence from YFP was detected in a single channel (channel 0) using an APD (Perkin Elmer model SPCM-AQRH-15).

Data acquisition uses the timetagger's GUI interface, which can be downloaded and installed from the "timetagger tools" package in the supplemental materials. Once launched by clicking on the "Timetag UI" icon, the GUI permits a user to set the data acquisition time, or to stop and start acquisition manually. The GUI also provides a monitor of detector count rates and continuously updated correlation functions during the course of an experiment. Data are stored as discussed above and in Fig. 3.

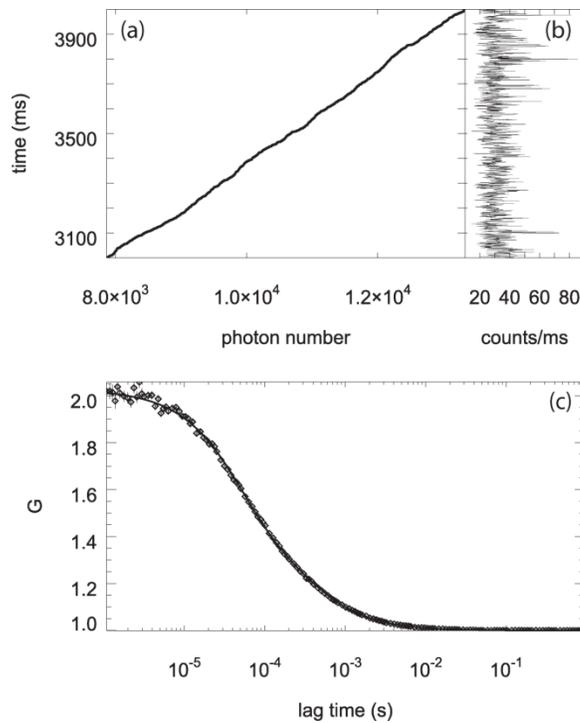

FIG. 9. Example of single-channel data. The sample is YFP at 5 nmol/L as described in the text. (a) A small segment of a 3.2 million photon data set, showing photon arrival time vs. photon number for photons arriving between 3 s and 4 s after the start of acquisition. (b) A histogram of the time-tag data shown in (a), showing the number of photons arriving in each 1 ms bin. (c) Diamonds: correlation function and uncertainties calculated from the full data set. The error bars, which except at short lag times are smaller than the symbols, are calculated using Eq. (4). The line is the result of a fit using Eq. (6); best fit parameters are $n$=1.68 ± 0.02, $\tau_D$ =209 ± 3 μs, $\tau_{tr}$ =50 ± 1 μs, and F = 0.42 ± 0.01. The $\chi^2$ per degree of freedom for this fit was 0.94.

An example of typical data acquired using this system are shown in Fig. 9(a), where the arrival times of photons at the channel 0 detector are plotted against photon number for a typical one second of data. A histogram of the time tags in 1 ms bins results in a more familiar plot of intensity (in photon counts/ms) vs. time shown plotted on its side in Fig. 9(b). Finally, as per the discussion in Sec. III, a correlation function is generated using the "fcs-corr" command in the "photon tools" package provided in the supplement [Fig. 9(c)].

As described in Ref. 8, correlation data can often be fit using a model that assumes free diffusion of the fluorophore:

$$G_D(\tau) = \frac{1}{n}\left[1 + \frac{\tau}{\tau_D}\right]^{-1}\left[1 + \frac{\tau}{w^2\tau_D}\right]^{-1/2} \tag{5}$$

Here $n$ is the average number of particles in the detection volume, $\tau_D$ is the diffusion time across the rms beam waist, and $w$ is the aspect ratio of the detection volume, which is approximated as a cylindrically symmetric three-dimensional Gaussian with $\langle \rho^2 \rangle = \langle z^2 \rangle / w^2$. However, since the correlation data shown here are calculated at shorter lags than possible in Ref. 8, we also account for the existence of a "dark" triplet state that gives rise to fluorescence fluctuations on much shorter time scales than particle diffusion. Assuming that transitions in and out of the triplet state do not affect diffusion, it is possible to separate out the diffusion and photophysical dynamics so that

$$G(\tau) = G_D(\tau)G_{tr}(\tau) . \tag{6}$$

where

$$G_{tr}(\tau) = \left(1 - F + F\exp\left(\frac{-\tau}{\tau_{tr}}\right)\right) \tag{7}$$

Here $F$ is the probability of finding the molecule in the triplet state at equilibrium in the excitation field, and $\tau_{tr}$ is a kinetic rate associated with the buildup of this population.[24] Using a calibration measurement as described in Sec. V.B. of Rieger *et al.*[8] we find $w=10.3\pm3$ and $\sqrt{\langle\rho^2\rangle}=278\pm13$ nm for our instrument.

To find the hydrodynamic radius of the protein, we fit Eq. (6) to the data shown in Fig. 9(c), with adjustable parameters $n$, $F$, $\tau_{tr}$ and $\tau_D$. For Brownian motion in two-dimensions, the diffusion time is given by $\tau_D=\langle\rho^2\rangle\big/4D$, from which $D$ can be determined. Then, using Eq. (1), and propagating the uncertainties in $w$ and $\tau_D$, we find $R_h=k_BT\big/6\pi\eta D=2.29\pm0.12$ nm. This is in agreement with the previously reported value $R_h=2.30\pm0.05$ nm for closely related Green Fluorescent Protein (GFP).[25] To the best of our knowledge, no other measurements of $R_h$ for YFP have been reported.

For comparison, Fig. 10 demonstrates typical data taken for protein aggregates. In this case, the diffusion across the sample volume is slow, and because there are multiple fluorophores in the protein particles, the signal is much brighter. Individual particles can be detected as large bright spikes in Fig. 10(b). It is immediately apparent from the correlation f unction, shown in Fig. 10(c) with the YFP monomer correlation function, that diffusion time for these aggregates, with their larger values of $R_h$, is much larger than for YFP. Additional information regarding this aggregated sample is given in the Supplementary Materials.

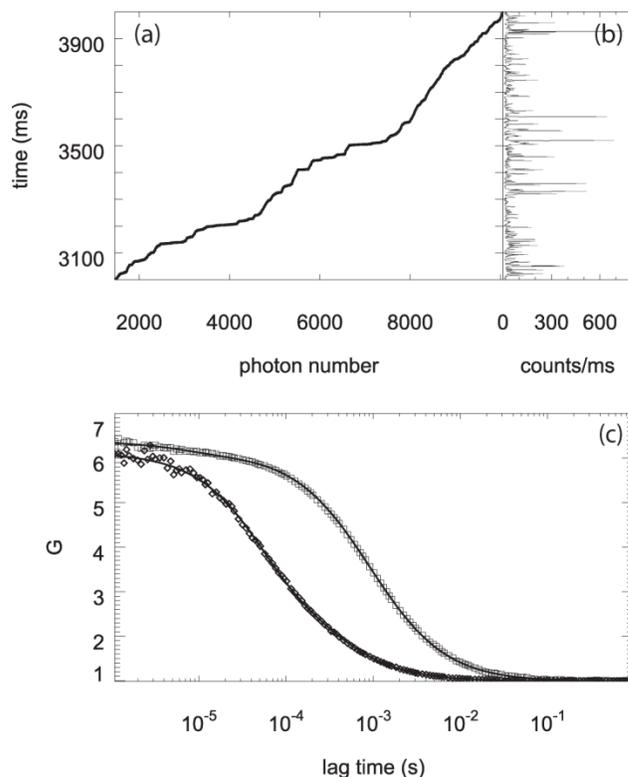

FIG. 10. Aggregated protein sample. (a) A small segment of a 2.8 million photon data set, showing photon arrival time vs. photon number for photons arriving between 3 s and 4 s after the start of acquisition. (b) A histogram of the time-tag data shown in (a), showing the number of photons arriving in each 1 ms bin. Here the prominent spikes correspond to individual, very bright, protein clumps diffusing through the detection volume. (c) Squares: correlation function calculated from aggregated protein sample, a portion of which is shown in (a) and (b). Triangles: YFP correlation function from Fig. 9(c) scaled up for comparison with aggregated protein. Lines are corresponding fits of Eq. (6) to the data.

**B. Dual-channel data demonstrating FRET.** An example of two-channel data acquired with this system is shown in Fig. 11. Here the sample is double-stranded DNA that is doubly-labeled with a pair of dyes, Cyanine 3 (Cy3) and Cyanine 5 (Cy5) suitable for demonstrating fluorescence resonance energy transfer (FRET). The DNA is diluted to 6.6 nmol/L in 20 mmol/L Hepes-NaOH (pH=7.5) and 50 mmol/L NaCl. Additional sample details can be found in the Supplementary Materials.

FRET describes the non-radiative energy transfer that can occur between an excited dipole (the donor, Cy3 here) and another dipole (the acceptor, Cy5 here) that has an absorption spectrum that overlaps the emission spectrum of the donor.[26] The probability of energy transfer, also called the FRET efficiency ($E$), is dependent on the distance between two dipoles, $R$, and can be expressed as

$$E = \left[ 1 + \left( R/R_0 \right)^6 \right]^{-1} \qquad . \qquad (8)$$

The Förster distance, $R_0$, can be calculated from the spectral properties of the donor and acceptor and their relative orientation.[26-28] A typical value of $R_0$ for freely rotating dyes is 5 nm, making FRET a sensitive "spectroscopic ruler"[27] in a small range around $R_0$. In two-channel FCS, the presence of FRET is evidenced by anticorrelation between the donor and acceptor signals that suppresses the amplitude of the cross-correlation: high FRET means the donor signal is suppressed in the presence of the acceptor.

For the data shown in Fig. 11, the donor was excited with the 514 nm line of the Argon-Krypton laser. Excitation power before entering the microscope was nominally 10 μW. Fluorescence from Cy3 and Cy5 were detected in separate channels using a dichroic mirror between the two detectors in Fig. 1. Data were acquired for 10 minutes; more than 4.1 million photons in the acceptor channel and 1.6 million photons in the donor channel were collected.

Figure 11(a) shows a typical 1 second of the data, histogrammed in 1 ms bins. In Fig. 11(b) are shown the auto- and cross-correlation functions for the entire data set. The donor channel autocorrelation is shown as triangles, and the acceptor channel as diamonds. In the absence of the acceptor, unavoidable cross talk from the Cy3 channel into the Cy5 channel would result in a correlated signal between the channels. In that case, auto- and cross-correlations all have approximately the same amplitude, reflecting primarily the concentration of the Cy3. In Fig. 11(b), the presence of a FRET pair results in an anti-correlated signal that shows up as a substantially suppressed cross-correlation (squares). The cross-correlation is also free of the large detector afterpulsing peak that dominates the autocorrelations for $\tau < 2$ μs.

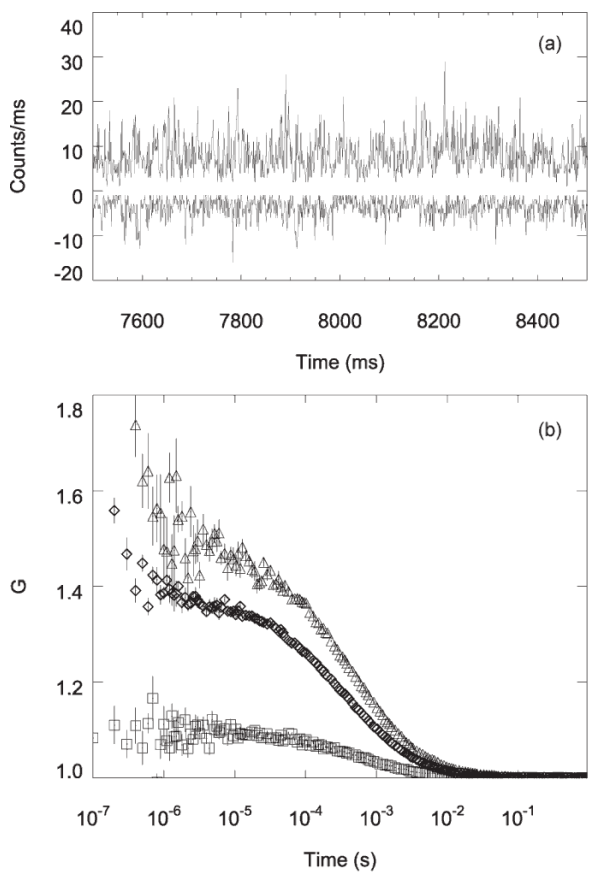

FIG. 11. Example of dual-channel data. The sample is DNA oligo at 6.6 nmol/L as described in the text. (a) A histogram of the time-tag data, showing the number of photons arriving in each 1 ms bin in a 1 s interval. For clarity, both channels are offset slightly from zero, and the donor channel is plotted inverted. (c) Diamonds: acceptor channel correlation function. Triangles: donor channel correlation function. Squares: cross correlation function. The error bars are calculated as described from Eq. (4).

**C. Alternating Laser Excitation and single-molecule FRET.** A more powerful method to measure FRET requires the use of using alternating laser excitation (ALEX).[29] In this example, unlike the previous two, the AOTF is in use and the sequencer records both strobe and delta input events. Single-molecule sensitivity permits the researcher to discern more accurate FRET values and distinguish different subpopulations of molecules.

ALEX is demonstrated in Fig. 12 using the same doubly-labeled DNA as in the previous section, except that the sample is diluted to the point where there is never more than one molecule in the detection volume at a time (50 pmol/L, see Supplementary Materials). Cy3 was excited with 514 nm at 205 μW, and Cy5 with 647 nm at 42 μW. The excitation light was switched every 50 μs using the sequencer outputs to drive the AOTF shown in Fig. 1. Fluorescence from Cy3 (donor) and Cy5 (acceptor) were detected as described above in channels 0 and 1 respectively; however here we distinguish between fluorescence acquired with 514 nm excitation (donor excitation) and with 647 nm excitation (acceptor excitation). In Fig. 12(a-c), one second worth of time tags are histogrammed in 1 ms bins. The three separate traces represent (a) donor channel photons during donor excitation, (b) acceptor channel photons during donor excitation (from FRET), and (c) acceptor channel photons during acceptor excitation. Photon bursts representing single fluorescent molecules crossing the detection volume are evident in all three measurements. Recall that for each photon (strobe event) the detection channel is recorded by the Channel bits, while the excitation source is determined by the state of the Channel bits at the preceding delta event. A total of 663,656 photons were acquired in all three channels during the 465 seconds of this measurement. ALEX data sets typically contain as many or more delta events from the sequencer as strobe events from the detectors; the two are distinguished by the Record Type bit in the data word (Fig. 3).

The utility of ALEX becomes evident in Fig. 6(d), which is a scatter plot of the stoichiometry *vs.* proximity ratio.[29] The Stoichiometry is given by the ratio of intensities:

$$S = \left( I_D^A + I_D^D \right) \Big/ \left( I_D^A + I_D^D + I_A^A \right),$$ (4)

where the subscripts denote the excitation source and the superscripts denote the detection channel: $D$ indicates donor and $A$ indicates acceptor. The proximity ratio is defined as

$$P = \left( I_D^A \right) \Big/ \left( I_D^A + I_D^D \right).$$ (5)

It is related to $E$ of Eq. (8), but lacks the corrections for background, cross-talk between the two channels, relative quantum-yields/detector efficiency, and direct excitation of the acceptor that are necessary for a quantitative measurement[29] of $E$. In Fig. 12(c), $S$ and $P$ are calculated for all photon bursts of greater than 30 photons; the intensities $I$ here represent the number of photons in any particular burst. We use a Bayesian inference algorithm to find photon bursts directly from time tags,[30] but a simpler thresholding scheme can also be used. The advantage of this two-dimensional representation of the data is that it permits immediate identification of peaks that are associated with a missing or photobleached donor or acceptor dye. Values of $S$ near 1 indicate a missing acceptor (donor-only peak), while very low values of $S$ correspond to a missing donor (acceptor-only peak). By plotting the data in this way we immediately see that the low value of $P$ with a peak near 0.15 is an artifact of acceptor photobleaching, and not associated with a configuration of the molecule. Figure 6(e) is a histogram of the proximity ratio regardless of $S$; the broad, high $P$ peak is what is expected for this flexible oligo.

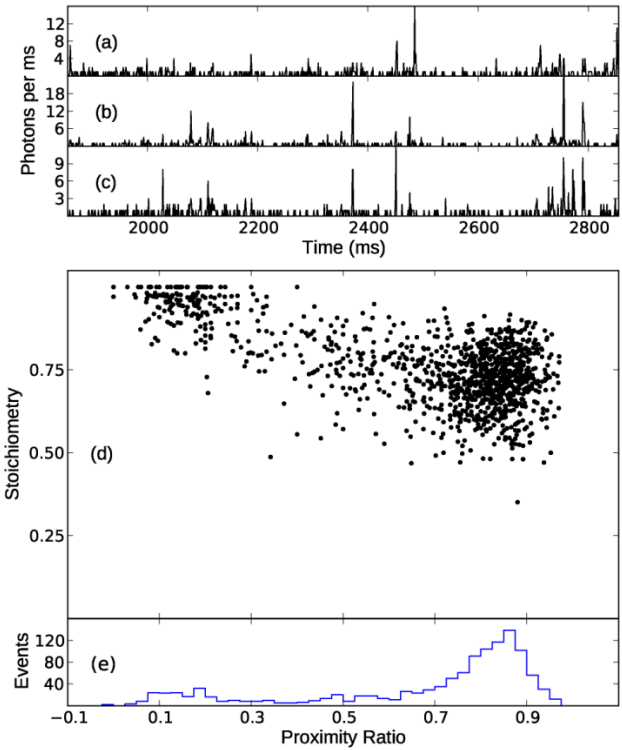

FIG. 12. Demonstration of ALEX using this system. (a), (b) and (c) are histograms of the time-tag data, showing the number of photons arriving in each 1 ms bin in a 1 s interval. The peaks correspond to individual molecules crossing the detection volume of the microscope. (a) fluorescence in the donor channel with the donor excitation. (b) fluorescence in the acceptor channel with donor excitation. (c) fluorescence in the acceptor channel with acceptor excitation. (d) is a scatter plot of stoichiometry vs. proximity ratio, as defined in the text. (e) is a proximity ratio histogram. The large peak on the right corresponds to FRET; the smaller peak on the left represents molecules with dark or photobleached acceptors.

## V. DISCUSSION

A wide variety of fluorescence techniques are now available for studying the organization, structure and structural changes of molecules both in living systems and *in vitro*. Among these, single-molecule sensitive FCS and FRET, and variations thereof, are arguably the most widely used. While both techniques have been around for many years, the advent and wide dissemination of single-molecule-sensitive optical instrumentation has ushered in a new era in molecular biophysics. Here we have presented a simple and inexpensive set of electronics and open-source software that can bring state-of-the-art capabilities to both the teaching and research laboratories. By providing outputs for laser control, this system makes it possible to run experiments with multiple dyes in several modalities without any physical changes to the apparatus or reprogramming required. This adds substantially to the information content of the data and simplifies experimental design and control.

In addition to the measurements demonstrated here, we note that this system can equally well be used in an undergraduate laboratory to measure the kinetics of intermolecular interactions,[8] perform time-correlated single-photon counting experiments relevant to the fundamentals of quantum mechanics,[31] and study phase transitions or measure particle size via dynamic light scattering.[32]

Because it accommodates multiple input and output channels, the electronics described here is broadly useful in the research laboratory. For example, three- and four-color FRET[33] with multicolor ALEX[34] and coincidence measurements with multiple dyes are straightforward to implement, as is the use of photoactivation. By adding appropriate electro-optic elements, the electronics can be used to switch polarization instead of wavelength, facilitating polarimetric measurements such as are used in rotational studies of single molecules.

 The use of an FPGA provides flexibility: with some modification, control channels might also be used to run a scanner for confocal imaging applications. High-end TCSPC systems typically include one or more time-to-digital converters (TDCs) with time resolution as small as a few picoseconds, suitable for applications such as fluorescence lifetime or polarization anisotropy lifetime and demonstrations of antibunching. Schemes for an FPGA implementation of a TDC[35] can now achieve 6 ps resolution[36] and would be the next obvious modification to this system.

Documentation, open-source software tools and firmware code for the FPGA can be found in the Supplementary Materials or downloaded from http://goldnerlab.physics.umass.edu/wiki/FpgaTimeTagger.

## VI. ACKNOWLEDGEMENTS

This work was supported by NSF grant MCB-0920139. We thank Dr. Daugherty for the gift of plasmid pYpet-His. Work on the WNV MTase was supported by grant AI07079201A1 and AI09433501 from NIH.

# I. HARDWARE CONSTRUCTION AND SPECIFICATIONS

Detailed construction, specifications, and parts list can be found in **tutorial.pdf** provided in the zipped archive. A brief overview is given here.

The time-tagger is built around a Xylo EM development board. This board provides a Altera Cyclone Field-Programmable Gate Array (FPGA) as well as a device (the Cypress FX2) for interfacing with a traditional USB bus. The FPGA consists of a set of logic elements (e.g. boolean logic gates), memories, and buses which can be connected to form complex, high-speed digital devices.

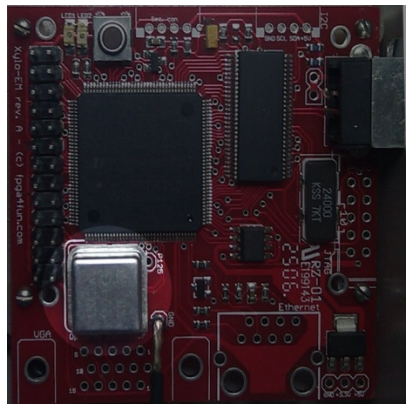

FIG. S1. Xylo EM development board with external oscillator highlighted.

The difficult work of supporting the FPGA logic is done by the Xylo board;[1] the remaining work consists primarily of making an enclosure for external connections and downloading the firmware. To keep out electromagnetic interference, the board was mounted in a metal box 6" on a side. The box was prepared with a hole on one side for the USB connector, and eight BNC connectors on the opposite side for outputs.

A standard half-size four-pin crystal oscillator (Digikey part number 535-9195-5-ND) was soldered in the space provided on the Xylo board (See Fig. S1). This location can be found adjacent to the FPGA and is marked by white silkscreen outlining the crystal package. The clock rate was derived from this oscillator but is multiplied by four onboard using a phase-locked loop. This functionality is configured by the firmware.

Stand-offs and bolts were then used to mount the board in the box with the on-board USB connector protruding through the hole on the back of the box. Eight BNC connectors were mounted on the front of the box. These include:

- Photon timing input channels 1-4, which are wired to pins 74,69,70,73 respectively;
- Delta timing input channels that are internally configured by the firmware and do not need external connections;
- Output control channels 1-4, which are wired to pins 75, 79, 97, 92, respectively.

Once assembled the unit was connected to a computer running a Microsoft Windows operating system to load the firmware using the software provided with the Xylo board.

The maximum event rate (*e.g.*, the maximum rate at which photons can be recorded) supported by the Tagger is determined by several factors. The greatest limit is set by the clock frequency, typically 128 MHz here. The maximum steady-state count rate is set by the bandwidth of the USB bus at 600,000 events/second. However, high count rate photon bursts can be absorbed by on-board buffers, allowing short bursts of events (e.g., photons) at the clock rate in each channel. In the case that a delta and a strobe event occur in the same clock cycle, the strobe event is dropped. The maximum burst length is roughly 2000 events, dictated by the size of the various buffers. In the event of a buffer overrun, a dropped data bit (bit 48) is set in the next successfully registered record to alert the user of lost data.

## II. SAMPLE PREPARATION

**Rhodamine 110 for calibration of the instrument.** Calibration of the instrument to find $\langle \rho^2 \rangle$ and $w$ was accomplished using Rhodamine 110 dye at 1 nM in the same buffer used to measure YFP. The known value of the diffusivity, $D$, was taken from Ref. 2. The calibration was done as described in Ref. 3.

**Preparation of protein aggregates.** The aggregated protein sample consists of West-Nile Virus Methyltransferase (WNV MTase) was expressed and purified as an N-terminal yellow fluorescent protein (YFP)-fusion protein. WNV MTase is a 35 kD globular protein; with YFP the total weight is 62.4 kD. Cloning, expression, and purification of the fusion protein was accomplished as follows. A DNA fragment representing the WNV NS5 MTase domain (N-terminal 300-amino acid) was PCR-amplified from a MTase construct[4] and inserted into a modified plasmid pYpet-His[5] at Bam HI and Not I sites. The construct was verified by DNA sequencing. *E.coli* strain MC1061 was used for protein expression. Cells were grown at 37ºC to an absorbance of 0.4 at 600 nm, then induced with 0.2% L-(+)-arabinose at 25 °C, and harvested by centrifugation. After sonication of the bacteria, soluble MTase containing an N-terminal YFP-His-tag were purified through a Ni-NTA column with an elution buffer containing 200 mM imidazole. The proteins (>90% purity) were further purified through a gel-filtration 16/60 Superdex column (Amersham), analyzed by SDS-PAGE, and quantified by Bradford protein assay (Bio-Rad).

To remove large protein aggregates, the sample was centrifuged at 13.2 krpm for 30 minutes at 4°C before dilution to a final concentration of 5.6 nmol/L protein in 50 mmol/L glycine (pH =10) and 0.16 mmol/L Tween 20 detergent. Deionized water used in sample preparation was treated with diethyl pyrocarbonate (DEPC). To prevent the protein from sticking, sample wells (Labtek Chambered Coverglass wells part number 155411) were incubated for 30 minutes in 10 mg/mL bovine serum albumin (BSA) solution, and then rinsed with DEPC treated deionized water before use.

**Preparation of doubly-labeled DNA for two-channel detection.** Double-stranded DNA with a single-stranded poly-dT region[6] consists of  OligoA: 5'-Cy5-GCCTCGCTGCCGTCGCCA-3'-Biotin hybridized to OligoD: 5'-TGGCGACGGCAGCGAGGCTTTTTTTTTTTTT-Cy3-T-3' at 6.6 nmol/L in 20 mmol/L Hepes-NaOH (pH=7.5) and 50 mmol/L NaCl.[6] DEPC treated deionized water was used in the sample preparation.  DNA with FRET pairs can be ordered from various companies (samples from Integrated DNA Technologies were used in the present work).

**For the ALEX data**, which is single-molecule sensitive, the sample was prepared at a concentration of 50 pmol/L in 20 mmol/L Hepes-NaOH (pH=7.8) and contains 0.0001 g/mL Tween (detergent) and 10% (v/v) glycerol.